# Impact of user distribution on optical wireless systems

Khulood D. Alazwary[1], Osama Zwaid Alsulami[1], Sarah O. M. Saeed[1], Sanaa Hamid Mohamed[1],
T. E. H. El-Gorashi[1], Mohammed T. Alresheedi[2] and Jaafar M. H. Elmirghani[1]
[1]School of Electronic and Electrical Engineering, University of Leeds, LS2 9JT, United Kingdom
[2]Department of Electrical Engineering, King Saud University, Riyadh, Kingdom of Saudi Arabia
elkal@Leeds.ac.uk, ml15ozma@leeds.ac.uk, elsoms@leeds.ac.uk, elshm@leeds.ac.uk,
t.e.h.elgorashi@leeds.ac.uk, malresheedi@ksu.edu.sa, j.m.h.elmirghani@leeds.ac.uk

**ABSTRACT**
In this paper, we investigate the impact of user distribution on resource allocation in visible light communication (VLC) systems, using a wavelength division multiple access (WDMA) scheme. Two different room layouts are examined in this study. Three 10-user scenarios are considered, while an optical angle diversity receiver (ADR) with four faces is used. A mixed-integer linear programming (MILP) model is utilized to identify the optimum wavelengths and access point (AP) allocation in each scenario. The results show that a change in user distribution can affect the level of channel bandwidth and SINR. However, a uniform distribution of users in the room can provide a higher channel bandwidth as well as high SINR above the threshold (15.6 dB) for all users compared to clustered users, which is a scenario that has the lowest SINR with supported data rate above 3.2 Gbps.

## 1. INTRODUCTION

At present, there is a massive increase in connected devices simultaneously with a rapid increase in the use of Information and Communication Technology (ICT) systems. Indeed, Cisco has estimated that the number of connected devices will be in the billions 2021[1]. Accordingly, wireless communication network (WCN) need to fulfill user demands, such as the need to realize high data rates. However, high data rates may become a challenge when employing the radio frequency (RF) spectrum which has a limited channel capacity and thus limited transmission rates. Optical wireless communication (OWC) systems are a potential solution that can support and meet the enormous growing demand for high data rates due to its high capacity. In addition to these features, OWC systems have higher levels of security compared to radio frequency wireless systems as optical signals cannot penetrate walls and therefore remain in the environment in which they originate. Therefore, OWC systems have gained increased interest [2]–[8], and are being considered in sixth generation communication (6G) systems. Furthermore, many demonstrations have shown that Visible Light Communication (VLC) systems, as a type of OWC system, can achieve capacities up to 20 Gbps [4]–[13].

Moreover, many studies attempted to improve the communication links by employing a variety of adaptation techniques such as beam power adaptation, beam angle adaptation, and beam delay adaptation [3], [14]–[20]. In addition to the beam adaptation techniques, diversity techniques have been utilized, such as angle diversity receivers, to overcome directed and undirected interference [21]–[24]. Uplink OWC systems have been discussed and evaluated in [25], [26]; however, further investigations are needed for energy efficiency [27]. In multi-user OWC systems, many multiple access schemes have been proposed and investigated, such as code division multiple access (CDMA) schemes where every user has a special code for simultaneous communications, and to mitigate interference between users [28]. To tackle interference different schemes have been proposed based on different orthogonal resources such as time, wavelength, and code [13], [29]–[33]. Wavelength division multiple access (WDMA) is a promising solution that uses multiple-colours to support multi-user access in VLC systems, and to reduce interference.

This paper studies the impact of user distribution on wavelengths and access point (AP) allocation. Laser Diodes (LDs) with four colures red, yellow, green, and blue (RYGB) are used in the optical access point. The latter is used as indoor white illumination in addition to providing high data rate communication [34]. This work examined three users in different room layouts, namely a conference table and cocktail party table layouts. In addition, a 4 branch angle diversity receiver was utilized as an optical receiver. The optimum allocation of resource in terms of access points (APs) and wavelengths is identified based on maximizing the overall users' SINRs through Mixed Integer Linear Programming (MILP). The rest of this paper is organized as follows: Section 2 describes the system configuration, including the room, users' locations, transmitters and receiver configuration, while the simulation results are shown and discussed in Section 3 and the conclusions are stated in Section 4.

## 2. SYSTEM CONFIGURATION

In this work, we considered an empty room that has dimensions (length × width × height) of 8 m × 4 m × 3 m. The room has no doors or windows and has eight lighting units as shown in Figure 1. For the communication context, the channel bandwidth can be evaluated by employing a ray tracing algorithm following [35], [36].

Only the line of sight, first and second order reflections were included in this simulation due to the insignificant impact of higher order reflections on the received power. Therefore, room surfaces (ceiling, walls and floor) were subdivided into small identical area that serve as secondary transmitters that reflect the light rays in the form of a Lambertian pattern as shown in [5]. The element's size in each surface has a significant role in the resolution of the simulation results. When smaller element sizes are used, a higher resolution is achieved, however, this increases the computation time required for the simulation [36]. All the communication links in the two different room layouts operated above the communication floor (CF) as shown in Figure 1.

An angle diversity receiver (ADR) that consists of four faces similar to [29] is used. Each branch in the ADR has a photodetector with a narrow Field of View (FOV) and two angles: Azimuth (*Az*) and Elevation (*El*) to collect signals from different sectors of the room and reduce the interference. The user's distribution around a conference table (350 length × 120 width ) is selected based on a standard in [37]. Table 1 shows the overall simulation parameters of the room, transmitter and receiver.

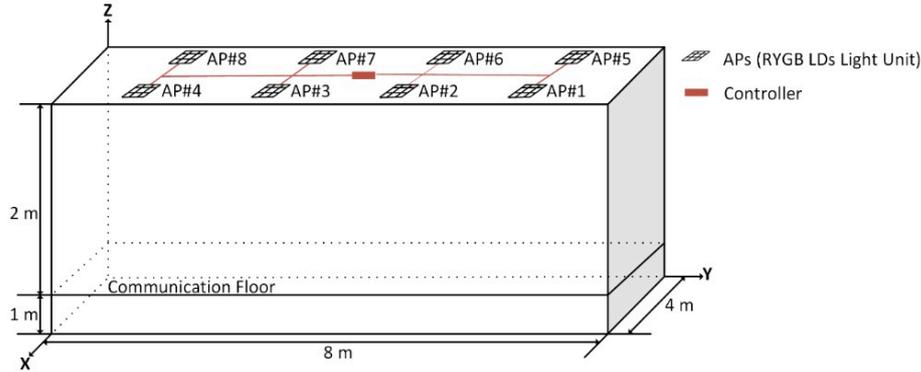

Figure 1: Room Configuration

**Table 1**. System Parameters

| Parameters | Configurations | |
|---|---|---|
| Walls and ceiling reflection coefficient | 0.8 | |
| Floor reflection coefficient | 0.3 | |
| Number of reflections | 1 | 2 |
| Area of reflection element | 5 cm × 5 cm | 20 cm × 20 cm |
| Order of Lambertian pattern, walls, floor and ceiling | 1 | |
| Semi-angle of reflection element at half power | 60° | |
| Number of RYGB LDs per unit | 12 | |
| Transmitted optical power of Red LD | 0.8 W | |
| Transmitted optical power of Yellow LD | 0.5 W | |
| Transmitted optical power of Green LD | 0.3 W | |
| Transmitted optical power of Blue LD | 0.3 W | |
| Total transmitted power of RYGB LD | 1.9 W | |
| **Room** | | |
| Width × Length × Height (x, y, z) | 4m × 8 m × 3 m | |
| Number of transmitters' units | 8 | |
| Transmitters locations (x, y, z) | (1 m, 1 m, 3 m), (1 m, 3 m, 3 m), (1 m, 5 m, 3 m), (1 m, 7 m, 3 m), (3 m, 1 m, 3 m), (3 m, 3 m, 3 m), (3 m, 5 m, 3 m) and (3 m, 7 m, 3 m) | |
| **Receiver** | | |
| Responsivity Red | 0.4 A/W | |
| Responsivity Yellow | 0.35 A/W | |
| Responsivity Green | 0.3 A/W | |
| Responsivity Blue | 0.2 A/W | |
| Number of photodetectors | 4 | |
| Area of the photodetector | 20 mm$^2$ | |

| Photodetector | 1 | 2 | 3 | 4 |
|---|---|---|---|---|
| Azimuth angles | 0° | 90° | 180° | 270° |
| Elevation angles | 60° | 60° | 60° | 60° |
| Field of view (FOV) | 25° | 25° | 25° | 25° |

| **Conference table** | |
|---|---|
| Receiver noise current spectral density | 4.47 pA/√Hz |
| Receiver bandwidth | 5 GHz |
| **Cocktail Party #1** | |
| Receiver noise current spectral density | 4.47 pA/√Hz |
| Receiver bandwidth | 5 GHz |
| **Cocktail Party #2** | |
| Receiver noise current spectral density | 4.47 pA/√Hz |
| Receiver bandwidth | 2.5 GHz |

## 3. SIMULATION SETUP AND RESULTS:

In this paper, we considered three different scenarios that represent three different user distributions in the same room configuration. In every scenario, a set of 10 users was distributed to produce a different setup. The first scenario is for users sitting around a (conference table) whereas the next two scenarios represent users distribution at (cocktail party#1) and (cocktail party#2). We considered two groups at every (cocktail party), and the distance between users in one group is 0.5m. Every group in (cocktail party#1) consists of 5 users, while the distribution of users in (cocktail party #2) among the two groups is 7 and 3 users. The 7-users group was placed at the room corner, whereas the 3-users group was placed in the middle of the room. In the (conference table) scenario, users were distributed around a conference table in the middle of the room where the distance between users is around 0.8 m as suggested in [37]. For each scenario, a wavelength division multiple access (WDMA) scheme is used to provide multiple access and a MILP model is utilized to optimize the resource allocation which maximizes the sum of all users' SINRs . Table 2 shows the users' locations and their resource allocation for every scenario. Moreover, the controller is placed on the ceiling of the room. The controller has prior knowledge of the users' locations (See Figure 1). Figure 2 shows an illustration of how the WDMA scheme works in VLC based on a scenario that consists of three APs, three users and two wavelengths (Red and Blue). The solid lines refer to links that have modulated data between a user and its assigned AP and wavelength, while the dashed lines refer to the interference between users using the same wavelength but different AP. Dotted lines indicate unmodulated wavelengths used for illumination only which is considered as background noise. As can be seen from Figure 2, User 1, who is the only one allocated to AP1 using the blue wavelength, is affected only by the background noise, while Users 2 and 3, who are allocated to AP2 and AP3 respectively using the same wavelength (Red), are affected by interference as well as background noise.

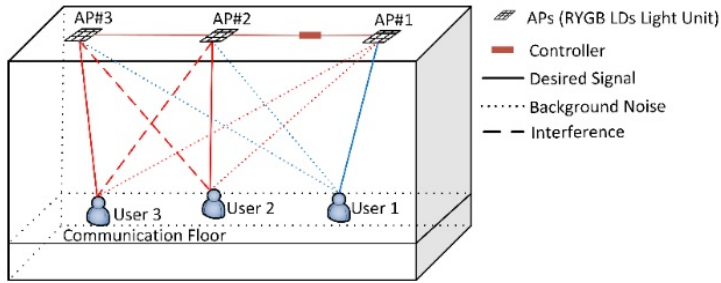

Figure 2: WDMA Example

**Table 2**.Scenarios with the optimized resource allocation.

| User | Conference table | | | | Cocktail Party #1 | | | | Cocktail Party #2 | | | |
|---|---|---|---|---|---|---|---|---|---|---|---|---|
| | Location (x, y, z) | AP | Branch | Wavelength | Location (x, y, z) | AP | Branch | Wavelength | Location (x, y, z) | AP | Branch | Wavelength |
| 1 | (1.5,2.5,1) | 1 | 4 | Red | (0.5,0.5,1) | 1 | 1 | Green | (0.5,0.5,1) | 1 | 1 | Blue |
| 2 | (1.5,3.5,1) | 2 | 3 | Red | (0.5,1.0,1) | 1 | 1 | Yellow | (0.5,1.0,1) | 1 | 1 | Yellow |
| 3 | (1.5,5.5,1) | 4 | 2 | Red | (0.5,1.5,1) | 2 | 2 | Yellow | (0.5,1.5,1) | 1 | 4 | Green |
| 4 | (1.5,4.5,1) | 3 | 3 | Red | (1.0,0.75,1) | 5 | 1 | Red | (0.5,2.0,1) | 2 | 2 | Yellow |
| 5 | (2.0,2.5,1) | 2 | 3 | Yellow | (1.0,1.25,1) | 1 | 4 | Red | (1.0,0.75,1) | 5 | 1 | Red |
| 6 | (2.0,5.5,1) | 3 | 3 | Yellow | (1.75,3.25,1) | 2 | 3 | Red | (1.0,1.25,1) | 1 | 4 | Red |
| 7 | (2.5,2.5,1) | 6 | 1 | Red | (1.75,3.75,1) | 6 | 1 | Yellow | (1.0,1.75,1) | 2 | 2 | Red |
| 8 | (2.5,3.5,1) | 6 | 1 | Yellow | (1.75,4.25,1) | 3 | 2 | Red | (1.75,3.75,1) | 6 | 1 | Red |
| 9 | (2.5,5.5,1) | 7 | 1 | Red | (2.25,3.5,1) | 6 | 1 | Red | (1.75,4.25,1) | 3 | 2 | Red |
| 10 | (2.5,4.5,1) | 7 | 1 | Yellow | (2.25,4.0,1) | 7 | 2 | Red | (2.25,4.0,1) | 7 | 2 | Red |

The optimisation of the resource allocation (APs and wavelengths) to each user in every distribution in both rooms' layout is optimized by using the MILP model. As a further step, the optical channel bandwidth and the SINR were verified at a fixed data rate for each user in every distribution in both room layouts, as shown in Figures 3, 4.

The simulation results in Figure 3 present the optical channel bandwidth for the different user distributions in both rooms' layout. The results vary due to the difference in the user distribution and location. The channel bandwidth in the conference table distribution is approximately between 5.5 GHz to 11.5 GHz, and the highest values of optical bandwidth are subject to the user location. Therefore, users who are close to the corner have the benefit of the channel reflections with lower delay spread.

In Figures 4 and 5, the SINR was evaluated at a fixed data rate of 7.1 Gbps for each user in the first two scenarios (conference table, cocktail party#1) except user 5 in (cocktail party#1) that has a data rate 5.4 Gbps due to the limited channel bandwidth associated with its location. Also, this exception includes users 1, 3, and 7 in (cocktail party#1) since they have SINRs lower than the threshold (15.6 dB) required for bit error rate of $10^{-9}$ when using On Off Keying (OOK) modulation. Thus, forward error correction (FEC) can be used to provide the same performance of (15.6 dB), similar to [29]. The last case applies to user 1 in (cocktail party#2), where its data rate decrease to 3.2 Gbps.

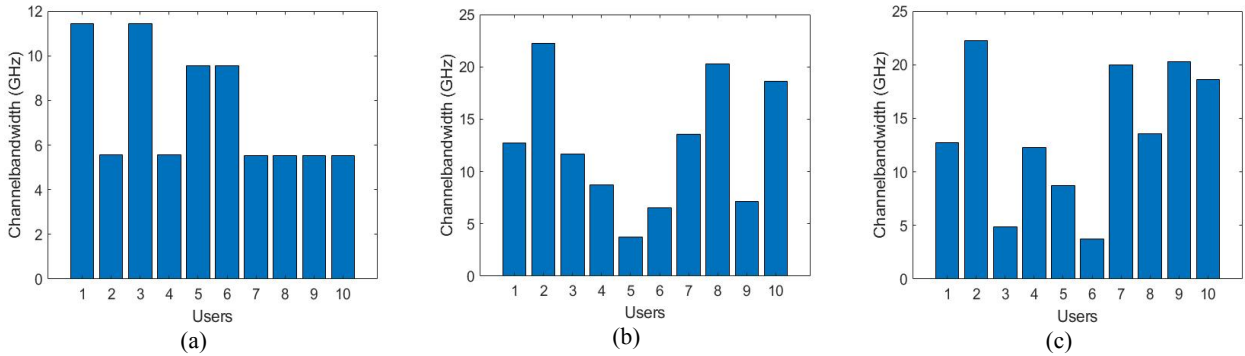

Figure 3: Channel bandwidth, (a) Conference table, (b) Cocktail Party #1, (c) Cocktail Party #2

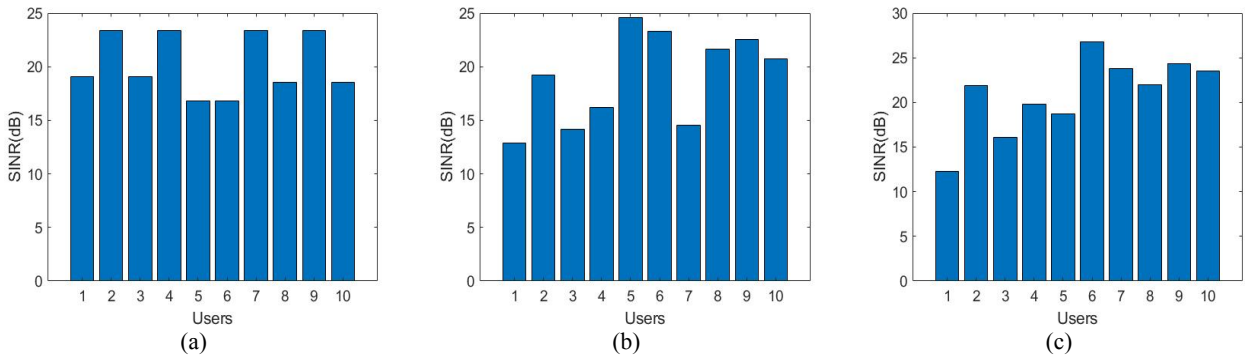

Figure 4: SINR, (a) Conference table, (b) Cocktail Party #1, (c) Cocktail Party #2

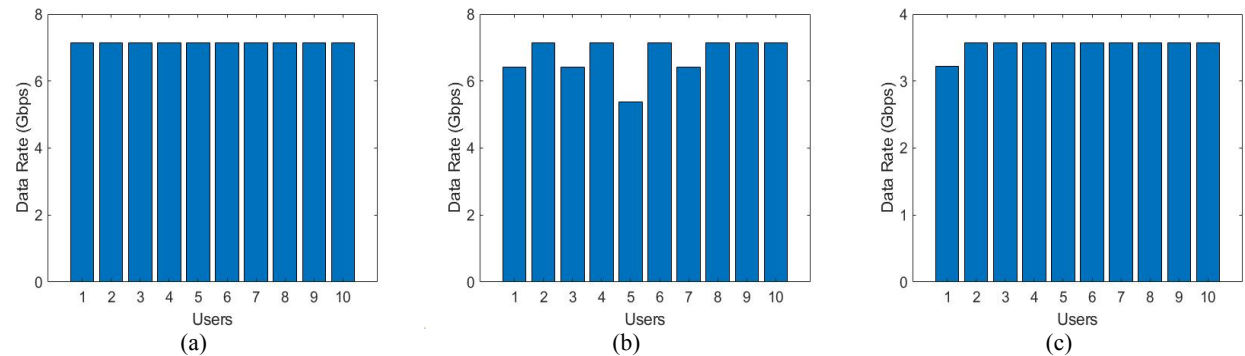

Figure 5: Data rate, (a) Conference table, (b) Cocktail Party #1, (c) Cocktail Party #2

## 4. CONCLUSIONS

This paper investigated the impact of user distribution on resource allocation in VLC systems using a proposed wavelength division multiple access (WDMA) scheme. Two different (rooms layout), and three different user clustering schemes were studied in this work. The setup had 10-users equipped with 4-branch ADR. A mixed-integer linear programming (MILP) model was utilized to identify the optimum resource allocation (AP and wavelength) where every AP consists of four wavelengths (red, green, yellow and blue). The results show that the optical channel bandwidth and SINR change when the user distribution changes and this affects the optimum allocation of wavelengths and APs to each user. All users in the first scenario (conference table) achieve a high SINR above the threshold (15.6 dB) compared to some users in the later scenarios (cocktail party#1 and cocktail party#2) that achieve an SINR less than the threshold, however all user locations can support high data rates above 3.2 Gbps.

## ACKNOWLEDGEMENTS

The authors would like to acknowledge funding from the Engineering and Physical Sciences Research Council (EPSRC) INTERNET (EP/H040536/1), STAR (EP/K016873/1) and TOWS (EP/S016570/1) projects. The authors extend their appreciation to the deanship of Scientific Research under the International Scientific Partnership Program ISPP at King Saud University, Kingdom of Saudi Arabia for funding this research work through ISPP#0093. OZA would like to thank Umm Al-Qura University in the Kingdom of Saudi Arabia for funding his PhD scholarship, KDA would like to thank King Abdulaziz University in the Kingdom of Saudi Arabia for funding her PhD scholarship, SOMS would like to thank the University of Leeds and the Higher Education Ministry in Sudan for funding her PhD scholarship. SHM would like to thank EPSRC for providing her Doctoral Training Award scholarship. All data are provided in full in the results section of this paper.

## REFERENCES

1. "The Cisco® Visual Networking Index (VNI) Global Mobile Data Traffic Forecast Update 2017 – 2021 - Wireless Broadband Alliance." [Online]. Available: https://wballiance.com/the-cisco-visual-networking-index-vni-global-mobile-data-traffic-forecast-update-2017-2021/. [Accessed: 28-Apr-2020].
2. A. T. Hussein, M. T. Alresheedi and J. M. H. Elmirghani, "Fast and Efficient Adaptation Techniques for Visible Light Communication Systems," *IEEE/OSA Journal of Optical Communications and Networking*, vol. 8, No. 6, pp. 382-397, 2016.
3. F. E. Alsaadi, M. A. Alhartomi, and J. M. H. Elmirghani, "Fast and efficient adaptation algorithms for multi-gigabit wireless infrared systems," *J. Light. Technol.*, vol. 31, no. 23, pp. 3735–3751, 2013.
4. A. T. Hussein and J. M. H. Elmirghani, "10 Gbps Mobile Visible Light Communication System Employing Angle Diversity, Imaging Receivers, and Relay Nodes," *J. Opt. Commun. Netw.*, vol. 7, no. 8, pp. 718-735, 2015.
5. A. T. Hussein and J. M. H. Elmirghani, "Mobile Multi-Gigabit Visible Light Communication System in Realistic Indoor Environment," *J. Light. Technol.*, vol. 33, no. 15, pp. 3293–3307, 2015.
6. M. T. Alresheedi and J. M. H. Elmirghani, "Hologram selection in realistic indoor optical wireless systems with angle diversity receivers," *IEEE/OSA Journal of Optical Communications and Networking,* vol. 7, No. 8, pp. 797-813, 2015..
7. A. T. Hussein, M. T. Alresheedi, and J. M. H. Elmirghani, "20 Gb/s Mobile Indoor Visible Light Communication System Employing Beam Steering and Computer Generated Holograms," *J. Light. Technol.*, vol. 33, no. 24, pp. 5242–5260, 2015.
8. A. G. Al-Ghamdi and J. M. H. Elmirghani, "Characterization of mobile spot diffusing optical wireless systems with receiver diversity," *ICC'04 IEEE International Conference on Communications,* vol. 1, pp. 133-138, Paris, 20-24 June 2004.
9. K.L. Sterckx, J.M.H. Elmirghani, and R.A. Cryan, "Sensitivity assessment of a three-segment pyrimadal fly-eye detector in a semi-disperse optical wireless communication link," *IEE Proceedings Optoelectronics*, vol. 147, No. 4, pp. 286-294, 2000.
10. A. Al-Ghamdi, and J.M.H. Elmirghani, "Line Strip Spot-diffusing Transmitter Configuration for Optical Wireless systems Influenced by Background Noise and Multipath Dispersion," *IEEE Transactions on communication,* vol. 52, No. 1, pp. 37-45, 2004.
11. A. G. Al-Ghamdi and J. M. H. Elmirghani, "Spot diffusing technique and angle diversity performance for high speed indoor diffuse infra-red wireless transmission," *IEE Proceedings Optoelectronics*, vol. 151, no. 1, pp. 46–52, 2004.
12. O. Z. Alsulami, M. O. I. Musa, M. T. Alresheedi, and J. M. H. Elmirghani, "Visible light optical data centre links," *Int. Conf. Transparent Opt. Networks*, vol. 2019-July, pp. 1–5, 2019.
13. A. Al-Ghamdi, and J.M.H. Elmirghani, "Analysis of diffuse optical wireless channels employing spot diffusing techniques, diversity receivers, and combining schemes," *IEEE Transactions on communication*, Vol. 52, No. 10, pp. 1622-1631, 2004.
14. F. E. Alsaadi and J. M. H. Elmirghani, "Adaptive mobile line strip multibeam MC-CDMA optical wireless system employing imaging detection in a real indoor environment," *IEEE J. Sel. Areas Commun.*, vol. 27, no. 9, pp. 1663–1675, 2009.
15. F. E. A. and J. M. H. Elmirghani, "High-speed spot diffusing mobile optical wireless system employing beam angle and power adaptation and imaging receivers," *J. Light. Technol.*, vol. 28, no. 6, pp. 2191–2206, 2010.
16. F. E. Alsaadi and J. M. H. Elmirghani, "Mobile multigigabit indoor optical wireless systems employing multibeam power adaptation and imaging diversity receivers," *J. Opt. Commun. Netw.*, vol. 3, no. 1, pp. 27–39, 2011.
17. F. E. Alsaadi and J. M. H. Elmirghani, "Performance evaluation of 2.5 Gbit/s and 5 Gbit/s optical wireless systems employing a two dimensional adaptive beam clustering method and imaging diversity detection," *IEEE J. Sel. Areas Commun.*, vol. 27, no. 8, pp. 1507–1519, 2009.
18. F. E. Alsaadi, M. Nikkar, and J. M. H. Elmirghani, "Adaptive mobile optical wireless systems employing a beam clustering method, diversity detection, and relay nodes," *IEEE Trans. Commun.*, vol. 58, no. 3, pp. 869–879, 2010.
19. M. T. Alresheedi and J. M. H. Elmirghani, "Performance Evaluation of 5 Gbit/s and 10 Gbit/s Mobile Optical Wireless Systems Employing Beam Angle and Power Adaptation with Diversity Receivers," *IEEE J. Sel. Areas Commun.*, vol. 29, no. 6, pp. 1328–1340, 2011.
20. M. T. Alresheedi and J. M. H. Elmirghani, "10 Gb/s indoor optical wireless systems employing beam delay, power, and angle adaptation methods with imaging detection," *J. Light. Technol.*, vol. 30, no. 12, pp. 1843–1856, 2012.

21. K. L. Sterckx, J. M. H. Elmirghani, and R. A. Cryan, "Pyramidal fly-eye detection antenna for optical wireless systems," in *IEE Colloquium on Optical Wireless Communications (Ref. No. 1999/128)*, 1999, pp. 5/1-5/6.
22. A. Al-Ghamdi and J. M. H. Elmirghani, "Optimization of a triangular PFDR antenna in a fully diffuse OW system influenced by background noise and multipath propagation," *IEEE Trans. Commun.*, vol. 51, no. 12, pp. 2103–2114, 2003.
23. H.-H. Chan, K. L. Sterckx, J. M. H. Elmirghani, and R. A. Cryan, "Performance of optical wireless OOK and PPM systems under the constraints of ambient noise and multipath dispersion," *IEEE Commun. Mag.*, vol. 36, no. 12, pp. 83–87, 1998.
24. A. G. Al-Ghamdi and J. M. H. Elmirghani, "Performance evaluation of a triangular pyramidal fly-eye diversity detector for optical wireless communications," *IEEE Commun. Mag.*, vol. 41, no. 3, pp. 80–86, 2003.
25. M. T. Alresheedi, A. T. Hussein, and J. M. H. Elmirghani, "Uplink design in VLC systems with IR sources and beam steering," *IET Commun.*, vol. 11, no. 3, pp. 311–317, 2017.
26. O. Z. Alsulami, M. T. Alresheedi, and J. M. H. Elmirghani, "Infrared Uplink Design for Visible Light Communication (VLC) Systems with Beam Steering," in *2019 IEEE International Conference on Computational Science and Engineering (CSE) and IEEE International Conference on Embedded and Ubiquitous Computing (EUC)*, 2019, pp. 57–60.
27. J. M. H. Elmirghani *et al.*, "GreenTouch GreenMeter core network energy-efficiency improvement measures and optimization," *IEEE/OSA J. Opt. Commun. Netw.*, vol. 10, no. 2, pp. A250–A269, 2018.
28. F. E. Alsaadi and J. M. H. Elmirghani, "Adaptive mobile spot diffusing angle diversity MC-CDMA optical wireless system in a real indoor environment," *IEEE Trans. Wirel. Commun.*, vol. 8, no. 5, pp. 2187–2192, 2009.
29. O. Z. Alsulami *et al.*, "Optimum resource allocation in optical wireless systems with energy-efficient fog and cloud architectures," *Philos. Trans. R. Soc. A Math. Phys. Eng. Sci.*, vol. 378, no. 2169, pp. 1–11, 2020.
30. Y. Wang, Y. Wang, N. Chi, J. Yu, and H. Shang, "Demonstration of 575-Mb/s downlink and 225-Mb/s uplink bi-directional SCM-WDM visible light communication using RGB LED and phosphor-based LED," *Opt. Express*, vol. 21, no. 1, p. 1203, Jan. 2013.
31. F.-M. Wu, C.-T. Lin, C.-C. Wei, C.-W. Chen, Z.-Y. Chen, and H.-T. Huang, "3.22-Gb/s WDM visible light communication of a single RGB LED employing carrier-less amplitude and phase modulation," in *2013 Optical Fiber Communication Conference and Exposition and the National Fiber Optic Engineers Conference (OFC/NFOEC)*, 2013, pp. 1–3.
32. J.M.H. Elmirghani, and R.A. Cryan, "New PPM CDMA hybrid for indoor diffuse infrared channels," *Electron. Lett*, vol 30, No 20, pp. 1646-1647, 29 Sept. 1994.
33. T. A. Khan, M. Tahir, and A. Usman, "Visible light communication using wavelength division multiplexing for smart spaces," in *2012 IEEE Consumer Communications and Networking Conference (CCNC)*, 2012, pp. 230–234.
34. A. Neumann, J. J. Wierer, W. Davis, Y. Ohno, S. R. J. Brueck, and J. Y. Tsao, "Four-color laser white illuminant demonstrating high color-rendering quality," *Opt. Express*, vol. 19, no. S4, p. A982, Jul. 2011.
35. J. R. Barry, J. M. Kahn, W. J. Krause, E. A. Lee, and D. G. Messerschmitt, "Simulation of Multipath Impulse Response for Indoor Wireless Optical Channels," *IEEE J. Sel. Areas Commun.*, vol. 11, no. 3, pp. 367–379, 1993.
36. F. R. Gfeller and U. Bapst, "Wireless in-house data communication via diffuse infrared radiation," *Proc. IEEE*, vol. 67, no. 11, pp. 1474–1486, 1979.
37. Workspace, "Meeting Table Size And Seating Guide." [Online]. Available: https://workspace.ae/content/meeting-table-size-guide. [Accessed: 28-Apr-2020].